\RequirePackage{ifpdf}
\documentclass[11pt,a4paper]{article}
\pdfoutput = 1
\usepackage{jcappub}
\usepackage{mathrsfs}
\usepackage{dcolumn}
\usepackage{bm}
\usepackage{color}
\usepackage{ulem}

\providecommand{\beqa}{\begin{eqnarray}}
 
 \providecommand{\rm}{\mathrm}
\providecommand{\eeqa}{\end{eqnarray}}

\newcommand{\beq}{\begin{equation}}
\newcommand{\eeq}{\end{equation}}

\newcommand{\A}{\alpha}
\newcommand{\B}{\beta}
\newcommand{\G}{\gamma}
\def\T{{\scriptscriptstyle {\rm T}}}
\def\IR{{\scriptscriptstyle {\rm IR}}}
\def\UV{{\scriptscriptstyle {\rm UV}}}

\def\46{{\scriptscriptstyle {\rm 4-6}}}
\def\24{{\scriptscriptstyle {\rm 2-4}}}
\newcommand{\nst}{n_{\ast}}
\DeclareMathOperator{\sign}{sign}

\title{Non-Unitary Evolution in the \\ General Extended EFT of Inflation \&   \\  Excited Initial States}

\author[a,b]{Amjad Ashoorioon}

\affiliation[a]{\small School of Physics, Institute for Research in Fundamental Sciences (IPM),\\
P.O. Box 19395-5531, Tehran, Iran}

\affiliation[b]{\small Riemann Center for Geometry and Physics, Leibniz Universit\"at Hannover \\
Appelstra\ss{}e 2, 30167 Hannover, Germany}

\emailAdd{amjad@ipm.ir}

\abstract{I study the ``general'' case that arises in the Extended Effective Field Theory of Inflation (gEEFToI), in which the coefficients of the sixth order polynomial dispersion relation depend on the physical wavelength of the fluctuation mode, hence they are time-dependent. At arbitrarily short wavelengths the unitarity is lost for each mode.  Depending on the values of the gEEFToI parameters in the unitary gauge action, two scenarios can arise: in one, the coefficients of the polynomial become singular, flip signs at some physical wavelength and asymptote to a constant value as the wavelength of the mode is stretched to infinity. Starting from the WKB vacuum, the two-point function is essentially singular in the infinite IR limit. In the other case, the coefficients of the dispersion relation evolve monotonically from zero to a constant value in the infinite IR. In order to have a finite power spectrum starting from the vacuum in this case, the mode function has to be an eigensolution of the Confluent Heun (CH) equation, which leads to a very confined parameter space for gEEFToI. Finally, I look at a solution of the CH equation which is regular in the infinite IR limit and yields a finite power spectrum in either scenario. I demonstrate that this solution asymptotes to an excited state in past infinity in both cases. The result is interpreted in the light of the loss of unitarity for very small wavelengths. The outcome of such a non-unitary phase evolution should prepare each mode in the excited initial state that yields a finite two-point function for all the parameter space. This will be constraining of the new physics that UV completes such scenarios.}

\subheader{IPM/P-2018/029}

\begin{document}
\maketitle

\section{Introduction}
\label{Intro}

In the previous studies on the effect of new physics on inflationary observables, it was often suggested that the new physics appearing at energy scale $\Lambda_{\rm UV}$ in most of the cases can modify the power spectrum, even though such corrections could be quite small. The effect was often expressed in terms of 
$\left(\frac{H}{\Lambda_{\rm UV}}\right)^n$ , where $n$ was found to be one \cite{Martin:2000xs, Martin:2003kp, Danielsson:2002kx, Ashoorioon:2004vm, Ashoorioon:2004wd, Ashoorioon:2005ep} or two \cite{Kaloper:2002uj,Kempf:2001fa}, although \cite{Bozza:2003pr} claimed that the effect depends on the vacuum state obtained by the minimization of the Hamiltonian on the new physics hypersurface crossing \cite{Brown:1978ar}. Of course, if such a state coincides with what was called ``calm excited states" in \cite{Ashoorioon:2010xg}, the power spectrum is left intact, although it will ultimately manifest itself in the higher point functions. In the opposite limit, there could be excited initial states that modify the power spectrum by a very large modulation factor which needs to be considered when the amplitude of the power spectrum is matched with observation \cite{Ashoorioon:2013eia, Ashoorioon:2014nta}.  In all the above cases, the effect turned out to be finite.

One of the methods used to model the effect of new physics at very large momenta was the Modified Dispersion Relation (MDR) method \cite{Martin:2000xs, Martin:2003kp}. The scheme was originally motivated by the studies in condensed matter physics that suggested departure from Lorentzian dispersion relation beyond some momentum cutoff \cite{Unruh:1994je,Corley:1996ar}. It was shown in \cite{Ashoorioon:2017toq} that two formalisms of excited initial state and MDR could be related to each other and how a range of possibilities that arise within the excited state setup, appears in the formalism of MDR by changing the functional behavior and/or the coefficients of the dispersion relation. Although in the study of the new physics on inflationary observables, the MDRs were introduced in the setup of the cosmological perturbation theory through some ad hoc condensed matter procedure, with the advent of the Effective Field Theory of Inflation (EFToI) \cite{Cheung:2007st} and Ho\v{r}avaIt-Lifshitz gravity \cite{Horava:2009uw,Horava:2008ih}, such MDRs found a clear origin within the field of high energy physics. Spatial higher derivative terms naturally can arise in the EFToI and the coefficients of the corrections to the Lorentzian dispersion relation, will not be necessarily small as  simple promotion of such MDRs from the Minkowski space-time suggested. 

The EFToI allows for a unified picture of different single field inflationary scenarios, emphasizing on the fluctuations instead of the action for the inflaton. Time diffeomorphism is broken in the unitary gauge and the inflaton fluctuations are set to zero. Deviations from the slow-roll inflation is encoded in the operators built out of the variables that respect the remaining symmetries of the system, {\it i.e.} the spatial diffeomorphisms. The perturbations of the inflaton are then absorbed to the metric, which will have three degrees of freedom. Besides the two transverse degrees of freedom, there is a longitudinal component, a Goldstone boson $\pi$, whose action can be recovered with the St\"uckelberg technique. This is similar to what happens in spontaneously broken gauge theories, where the gauge field acquires a longitudinal component and becomes massive. The nonlinear sigma model describing the longitudinal components is UV completed to the Higgs theory. Here the nonlinear sigma model that describes the Goldstone boson is UV completed to the theory of inflaton with linear representation of time diffeomorphism.

The authors of EFToI ~\cite{Cheung:2007st} proved that the most general action for the single field inflation perturbations can be derived from implementing the St\"uckelberg procedure on the unitary gauge action which contains operators built out of any function of time, the $g^{00}$ component of the metric and the perturbations of the extrinsic curvature of constant inflaton hypersurfaces around the FRW background, $\delta K_{\mu\nu}$. The standard slow-roll inflation action can be recovered from the operators with time dependence and at most linear in terms of $g^{00}$ (mass dimension zero operators) in the decoupling limit. Higher dimensional operators are supposed to quantify the deviations from the slow-roll inflation. Various inflationary models, such as DBI inflation \cite{Alishahiha:2004eh}, with speed of sound different from one, or Ghost Inflation \cite{ArkaniHamed:2003uz, Gwyn:2012mw}, with leading spatial gradient term proportional to $(\nabla^2 \pi)^2$\cite{Cheung:2007st} can be reproduced in this way. In particular, for the class of DBI inflationary models, it is shown how reducing the speed of sound naturally enhances the non-gaussianity. In \cite{Cheung:2007st}, only up to mass dimension two operators were included in the unitary gauge action, which would lead to dispersion relations with quartic corrections to the Lorentzian dispersion relation. Although operator proportional to $(1+g^{00})^2$ modifies the speed of sound from one, in presence of  operators with mass dimension zero which is constructed out of $\delta K_{\mu\nu}^2$, one can again make the speed of sound equal to one. The nonlinear nature of time diffeomorphism will lead to non-vanishing higher point correlation function that can in principle be measured or bounded.

Although by including additional higher dimensional operators, one could modify the dispersion relation even further to obtain $\omega^2\sim k^{2n}$, with $n\geq 3$, \cite{Cheung:2007st} argued that these new terms would not be compatible with an  EFT formalism, since the interacting operators will become strong at low energy, which invalidates the EFT formalism. In \cite{Ashoorioon:2018uey}, nonetheless, we argued that such a dispersion relation is allowed when combined with lower order dispersion relations. As we showed in \cite{Ashoorioon:2018uey}, the terms in the unitary gauge action that leads to $k^6$ dispersion relation, naturally lead to $k^4$ and $k^2$ in the dispersion relation as well. In presence of such lower order dispersions, the the issue of strong coupling at low energy can be avoided. For that, the energy scale the dispersion relation has to change from $\omega^2\sim k^6$ to $\omega^2\sim k^4$, $\Lambda_{\rm dis}^\46$, has to be bigger than the strong coupling energy at low energy for pure $k^6$ dispersion relation, $\Lambda_6^\IR \ll \Lambda_{\rm dis}^\46$. In addition the scale $\Lambda_4$, where the $\omega^2\sim k^4$ theory becomes strongly coupled, should not be below $\Lambda_{\rm dis}^\46$. Also, the $\omega^2\sim k^2$ should not become strongly coupled at a scale $\Lambda_2$  below the scale $\Lambda_{\rm dis}^\24$ where the dispersion relation crosses to the quartic form, {\it i.e} $\Lambda_{\rm dis}^\24 \ll \Lambda_2$. The EFT formalism is viable on energy scales below ${\rm \Lambda}_{\rm UV}\equiv\min\{\Lambda_b, \Lambda_6^\UV\}$, where $\Lambda_b$ is the energy scale at which time diffeomorphism gets spontaneously broken by the inflaton background and $\Lambda_6^\UV$ is the UV strong coupling scale of the $\omega^2\sim k^6$ theory.

Out of the  operators up to mass dimension 4 that were added in the EEFToI, the mass dimension 3, which is  a parity-violating operator, led to only quadratic and quartic contributions to the dispersion relation but four operators with mass dimensions 4, which were built out of various indicial combinations of  $(\nabla K^{\mu}_{\ \nu})^2$ operators, led to sextic correction, in addition to the quadratic and quartic ones. However two of those operators, namely $(\nabla_{\mu} \delta K^{\nu\gamma})(\nabla^ {\mu} \delta K_{\nu\gamma})$ and $(\nabla_{\mu} \delta K^\nu_{\ \nu})^2$, contained Ostrogradski ghosts at the level of perturbations, either in the tensor and/or scalar sectors. The other two indicial combinations, namely $(\nabla_{\mu} \delta K^\mu_{\ \nu})(\nabla_{\gamma} \delta K^{\gamma\nu})$ and $\nabla^ {\mu}\delta K_{\nu\mu}\nabla^ {\nu}\delta K_{\sigma}^{\sigma}$, however, did not contain any ghost. In general it leads to sextic polynomial dispersion relation where the coefficients of the dispersion relation are dependent on the physical scale of the fluctuations. The scale dependence disappears for a specific sector in which certain relations between the coefficients of two healthy operators is satisfied and, in \cite{Ashoorioon:2018uey}, we focused on this sector of the theory. However in general the coefficient of the dispersion relation are dependent on the physical wavelength of the mode and hence are time-dependent. The analysis of the general case that can arise in the setup is the subject of the paper.

As we will see, in the general case, even if the parameters of the theory has been adjusted such that the Goldstone mode is decoupled from gravity at the horizon crossing, due to the time dependence of the coefficients of the dispersion relation, they have been coupled to gravity at sub-Planckian scales. Of course the EFT breaks down at those scales, because one would need to use the quantized theory of gravity, which is lacking. As I will point out, two scenarios are conceivable depending on the couplings of the operators in the unitary gauge action. In one, the coefficients of the dispersion relation become singular and change sign at some point throughout the evolution. In another one, the coefficients increase monotonically from zero as the modes are stretched. Starting from the positive frequency WKB vacuum, in the first case, the two point function never becomes constant, but evolves at superhorizon scales and becomes infinite in the infinite IR limit. For the other case, the two point function although becomes finite, in order for the mode function to  satisfy the Wronskian condition, one has to demand that the mode function is eigensolution of the differential equation and hence the parameters of the problem has to satisfy certain constraints, which is expressed in terms of continued fraction relations. In both two scenarios, the two point function is never finite for all the regions of parameter space. On the other hand imposing that the mode function would be regular in the infinite IR limit, we will show that the mode function is a combination of positive and negative frequency WKB modes, namely it is an excited state. In other words, in order to obtain sensible results for the inflationary two-point function for all the parameter space, one {\it has to} assume that the mode function is an excited state. We interpret this result  in the light of the mixing of the Goldstone mode with the gravity at infinitely short scales, which turns out to be sub-Planckian. Truncating the interacting regime, one has to replace the initial state of the theory with an excited states instead.

The outline of this paper is as follows. In Sec.~\ref{gEEFT}, I  briefly review the formalism of the EEFToI and how the absence of ghost at the level of perturbations in the scalar and tensor sector constraints the parameter space of the EEFTo1. In contrast to our previous study in \cite{Ashoorioon:2018uey}, I analyze the {\it general} case in the EEFToI in which the coefficients of the dispersion relation become time-dependent. Depending on the parameters in the the gEEFToI, two scenarios can arise. In one case, there will be a change of sign of the coefficients of the dispersion relation at a singular point throughout the evolution. In another one, the coefficients vary monotonously from  close to zero to their final constant value in infinite IR. We show that in the first case the two point function becomes singular in the infinite IR limit. In the second case, the power spectrum becomes finite only in certain regions of parameter space.  At the end, I demonstrate that the solution which is regular and finite in the infinite IR limit is actually an excited state in the the deep UV. I try to interpret this result in the light of violation of the unitarity at very small physical wavelengths.  Finally, I conclude the paper and outline directions for future research.

\section{General Extended EFT of inflation}
\label{gEEFT}

Let me begin by briefly reviewing the EEFToI framework which allows for the inclusion of sixth order corrections to the dispersion relation, without introducing ghosts, in the framework of the EFT of inflation. In the EFT of inflation, one restricts to  hypersurfaces with no inflaton fluctuations (the unitary gauge).  The most general action around the FRW background could be written out of the terms that respect the remaining 3-diffeomorphisms. This includes $g^{00}$, any pure functions of time $f(t)$, and terms which depend on the extrinsic curvature of the constant time hypersurfaces, $K_{\mu\nu}$. One can show that any other operator can be expressed in terms of the ones built out of these variables. The resulted Lagrangian around the flat FRW could be written as \cite{Cheung:2007st}
\beq
\mathcal{L}
=
M_{\rm Pl}^2
\left[
\frac12\, R
+\dot H\,  g^{00}
-\left(3\, H^2 +\dot H\right)
\right]
+\sum_{m\geq 2}\,
\mathcal{L}_m (g^{00}+1,\delta K_{\mu\nu}, \delta R_{\mu\nu\rho\sigma}, \nabla_\mu;t)
\ ,
\label{genspacei}
\eeq
where $\mathcal{L}_m$ represent functions of order $m$ in $g^{00}+1$, $\delta K_{\mu\nu}$ and $\delta R_{\mu\nu\rho\sigma}$. Furthermore, the pure time-dependent term and coefficient of $g^{00}$ have become functions of $H$ and $\dot{H}$ by satisfying the Friedmann equations in the FRW limit.

Once the generic action in unitary gauge has been written, the gauge condition can be relaxed by allowing the time transformation $t\to t+\xi^0(x^\mu)$. Since the action is no more restricted to a particular time slicing,  time diffeomorphism invariance has to be restored again. This can be achieved by substituting  $\xi_0(x^\mu)$ with a field $-\pi(x^\mu)$ and requiring that it shifts as $\pi(x^\mu)\rightarrow \pi(x^\mu)-\xi_0(x^\mu)$ under time diffeomorphisms.
Note that when we perform $t\to t+\xi^0(x^\mu)$, we no longer expect the perturbations in inflaton field, $\phi$,  to be zero. In fact, by introducing the Goldstone boson we are representing these perturbations.

This action is very complicated in general. However, the advantage of using this approach is that, in certain limits known as the decoupling limit, one can ignore all the metric perturbations in the action. It can be shown that if we implement this procedure for the action~\eqref{genspacei}, assuming $\mathcal{L}_m=0$ $\forall\, m$, we obtain
\beq
\mathcal{L}_{\rm slow-roll}= -M_{\rm Pl}^2\,  \dot{H}\left(\dot{\pi}^2-{(\partial \pi)^2\over a^2}\right).
\eeq
One can show, using proper gauge transformations, that $\pi$ is related to the conserved quantity $\zeta$ by $\zeta=-H\pi$. Substituting $\zeta$ in the above action reduces it to the standard slow-roll inflationary action for $\zeta$. Since we are interested in deviations from the standard slow-roll model, we will turn on the coefficients of the operators in $\mathcal{L}_m$. In the EEFToI, we extended the framework of EFToI to incorporate the mass dimension three and four operators in the unitary gauge action. Such operators would lead to sixth order corrections to the dispersion relation which the authors had excluded as they had argued that the pure $\omega^2\sim k^6$ dispersion relation would invalidate the EFT rationale as it leads to large interacting three point functions. However we argued that the exclusion of such operators are not justified for a mixed dispersion relation that deforms to $k^4$ and/or $k^2$ at low energies. In fact the new higher dimensional mass operators added in the unitary gauge, not only led to the sixth order dispersion relation, but also quartic and quadratic terms too. In this sense, the theory has a self-healing property.

Focusing only on the terms that contribute to the quadratic action of $\pi$ and can change the dispersion relation up to six order, the unitary gauge action is
\beq
\mathcal{L}_{\rm EEFToI} = \mathcal{L}_{\rm slow-roll} + \mathcal{L}_2\,,
\label{LEFT}
\eeq
where
\begin{eqnarray}
\label{eq:actiontad}
 \mathcal{L}_2
&=&
\frac{M_2^4}{2!}\,(g^{00}+1)^2
+\frac{\bar M_1^3} {2}(g^{00}+1) \delta K^\mu_{\ \mu} -\frac{\bar M_2^2}{2}\, (\delta K^\mu_{\ \mu})^2
-\frac{\bar M_3^2}{2}\, \delta K^\mu_{\ \nu}\,\delta K^\nu_{\ \mu}\nonumber\\
&&+\frac{\bar{M}_4}{2} \nabla^\mu g^{00}\nabla^\nu\delta K_{\mu \nu}-\frac{\delta_1}{2}\,(\nabla_{\mu} \delta K^{\nu\gamma})(\nabla^ {\mu} \delta K_{\nu\gamma})
-\frac{\delta_2}{2}\,(\nabla_{\mu} \delta K^\nu_{\ \nu})^2
\nonumber\\
&&-\frac{\delta_3}{2} \,(\nabla_{\mu} \delta K^\mu_{\ \nu})(\nabla_{\gamma} \delta K^{\gamma\nu})-\frac{\delta_4}{2} \,\nabla^ {\mu}\delta K_{\nu\mu}\nabla^ {\nu}\delta K_{\sigma}^{\sigma},
\end{eqnarray}
The first term, $(g^{00}+1)^2$, is the operator that modifies the speed of sound for scalar perturbations from the speed of light. Noting that $(g^{00}+1)$ has zero mass dimension, powers of $(g^{00}+1)^n$ with $n\geq 3$ could also be included which result in more general K-inflationary models. In this work, without loss of generality, we only include the quadratic term in $(g^{00}+1)$ as it only modifies the speed of sound for scalar perturbations. The mass dimension 1 term, $(g^{00}+1) \,\delta K^\mu_{\ \mu} $, is not symmetric under time reversal, and was already analysed in~\cite{Cheung:2007st,Bartolo:2010bj,Ashoorioon:2011eg}. $ \frac{\bar M_2^2}{2}\, (\delta K^\mu_{\ \mu})^2$ and $\frac{\bar M_3^2}{2}\,\delta K^\mu_{\ \nu}\,\delta K^\nu_{\ \mu}$ are the operators that lead to generalized Ghost Inflation with a quartic correction to the dispersion relation~\cite{ArkaniHamed:2003uz, Cheung:2007st}. Mass dimension three operator $\nabla^\mu g^{00}\nabla^\nu\delta K_{\mu \nu}$, as we will see, will lead to at most quartic corrections, but mass dimension four operators, $(\nabla K_{\mu\nu})^2$ operators will also generate the sixth order contribution to the dispersion relation. The reason advocated in~\cite{Cheung:2007st} for discarding these operators is that sixth order corrections to the dispersion relation would make higher derivative operators relevant, signaling the presence of an IR strong coupling regime. However, as explained in Sec.~\ref{Intro}, one can still have a sensible EFT description if the IR strong coupling scale of the $\omega^2\sim k^6$ theory is below the scale $\Lambda_{\rm dis}^\46$ where the dispersion relation becomes dominated by the quartic term.

Studying the tensor perturbations in the transverse traceless gauge,
\beq
g_{ij}=a^2 (\delta_{ij}+\gamma_{ij})\, ,
\eeq
where
\beq\label{tt-cond}
\gamma_{ii}=0\,,\qquad\quad \partial_{i}\gamma_{ij}=0\ .
\eeq
Only $-\frac{\bar M_3^2}{2}\, \delta K^\mu_{\ \nu}\,\delta K^\nu_{\ \mu}$, from the operators existed in the EFToI \cite{Cheung:2007st}, affects the equation of motion for $\gamma_{ij}$ by modifying the speed of propagation of gravitational waves,
\beq\label{M3bar-negative-sign}
c_\T^2=\left(1-\frac{\bar M_3^2}{M_{\rm Pl}^2}\right)^{-1}
\ .
\eeq
From the new operators in the EEFToI \cite{Ashoorioon:2018uey}, only $\frac{\delta_1}{2}\,(\nabla_{\mu} \delta K^{\nu\gamma})(\nabla^ {\mu} \delta K_{\nu\gamma})$ contributes to the action of tensor perturbations after the TT gauge is imposed. It actually contributes a ghost, which in order to get rid of, one has to set to zero the coefficient of the operator,
\beq
\delta_1=0
\ .
\eeq
As for the scalar sector, the Goldstone boson $\pi$, can be made explicit in the action through the St\"uckelberg method. Evaluating the action explicitly for $\pi$ in Fourier space in the limit $\dot{H}\to 0$, the EEFToI Lagrangian~\eqref{LEFT} in the unitary gauge leads to the following second order Lagrangian:
\begin{eqnarray}
\mathcal{L}_{\rm EEFToI}^{(\pi)}&=& M_p^2 \dot{H} (\partial_{\mu}\pi)^2+2 M_2^4 \dot{\pi}^2-\bar M_1^3 H\left(3\dot{\pi}^2-\frac{(\partial_i \pi)^2}{2 a^2}\right)-\frac{\bar M_2^2}{2} \left(9H^2 \dot{\pi}^2-3 H^2 \frac{(\partial_i \pi)^2}{a^2}\right.\nonumber\\&+&\left.\frac{(\partial_i^2 \pi)^2}{a^4} \right)
-\frac{\bar M_3^2}{2} \left(3H^2 \dot{\pi}^2-H^2 \frac{(\partial_i \pi)^2}{a^2}+\frac{(\partial_j^2 \pi)^2}{a^4}\right) \nonumber\\
&+& \frac{\bar M_4}{2}\left(\frac{k^4 H \pi^2}{a^4}+\frac{k^2 H^3 \pi^2}{a^2}-9 H^3\dot\pi^2\right)\nonumber\\
&-&\frac{1}{2} \delta_1 \left(\frac{k^6 \pi ^2}{a^6}-\frac{3 H^2 k^4 \pi ^2}{a^4}-\frac{k^4 \dot{\pi}^2}{a^4}+\frac{4 H^4 k^2 \pi ^2}{a^2}-6 H^4 \dot{\pi}^2-3 H^2 \ddot{\pi}^2\right)\nonumber\\
&-&\frac{1}{2} \delta_2 \left(\frac{k^6 \pi ^2}{a^6}+\frac{H^2 k^4 \pi ^2}{a^4}-\frac{k^4 \dot{\pi}^2}{a^4}+\frac{6 H^4 k^2 \pi ^2}{a^2}-9 H^2 \ddot{\pi}^2\right)\nonumber\\
&-&\frac{1}{2} \delta_3 \left(\frac{k^6 \pi ^2}{a^6}+\frac{3 H^2 k^4 \pi ^2}{a^4}+\frac{ H^2 k^2 {\dot{\pi}} ^2}{a^2}-9 H^4 \dot{\pi}^2\right)\nonumber\\ &-&\frac{1}{2}\delta_4 \left(\frac{k^6 \pi ^2}{a^6}+\frac{ H^2 k^4 \pi ^2}{2 a^4}+\frac{9 H^4 k^2 \pi ^2}{2 a^2}+\frac{3 H^2 k^2 \dot{\pi}^2}{a^2}+\frac{27}{2} H^4 \dot{\pi}^2\right)\,.\nonumber\\
\label{Lpi}
\end{eqnarray}
The mass dimension 4 operators, $-\frac{\delta_1}{2}(\nabla_{\mu} \delta K^{\nu\gamma})(\nabla^{\mu} \delta K_{\nu\gamma})$ and $-\frac{\delta_2}{2}(\nabla_{\mu} \delta K^\nu_{\ \nu})^2 $, yield $\ddot{\pi}^2$ after St\"uckelberg transformation and thus contain Ostrogradski's ghost~\cite{Ostrogradski}. Such a ghost term could be avoided if $\delta_1=-3\,\delta_2$, but as we needed $\delta_1=0$ to avoid ghost at the level of tensor perturbations, one has to assume that both $\delta_1$ and $\delta_2$ are simultaneously zero. The equation of motion for $u_k=a\,\pi_k$, in conformal time,  $d\tau\equiv {dt}/{a}$, and in the limit $\dot{H}\to 0$ is
\beqa
&&u_k''+ \frac{2 k^2  H^3 (\delta_3+3 \delta_4 )}{a A_1}\, u_k'+u_k \left(\frac{C_1}{A_1}\frac{k^6}{a^4}+\frac{D_1}{A_1}\frac{k^4}{a^2}+\frac{F_1}{A_1}k^2-\frac{a''}{a}\right)=0\,,
\eeqa
where
\beqa
&&A_1=-2 M_{\rm Pl}^2 \dot{H}+4 M_2^4-6\bar M_1^3 H-9 H^2 \bar M_2^2-3 H^2 \bar M_3^2+2 H^4 F_0(k,\tau)\,,\nonumber\\
&&C_1=\delta_3+\delta_4\,,\nonumber\\
&&D_1=  \bar M_2^2+\bar M_3^2+H^2 \frac{\delta_4}{2}+3H^2 \delta_3-\bar M_4 H\,,\nonumber\\
&& F_1=-2 M_{\rm Pl}^2 \dot{H}-\bar M_1^3 H-3H^2 \bar M_2^2-\bar M_3^2 H^2+3 H^4 \left(\delta_3+\frac{3}{2}\delta_4\right)-\bar{M}_4 H^3
\ ,
\eeqa
and $F_0(k,\tau)$ is defined as
\beq
F_0(k,\tau)=\frac{9}{2}\delta_3-\frac{27}{4}\delta_4- \frac{k^2}{2 a^2 H^2}(\delta_3+3\delta_4)-\frac{9}{2H}\bar M_4
\ ,
\eeq
where the mixing with gravity has been neglected. Noting that the canonical $\pi_c\sim \sqrt{A_1(k/aH)}\, \pi$ and $\delta g^{00}_{c}\sim M_{\rm Pl}\, \delta g^{00}$, the mixing momentum between gravity and the Goldstone mode $\pi$ can be neglected at energies
\beq
k> k_{\rm mix}\sim \frac{\sqrt{A_1(k/aH)}}{M_{\rm Pl}}
\ ,
\eeq
and the physics of the longitudinal components of the metric can be studied, at sufficiently high momenta/energies, concentrating on the scalar Goldstone mode. This is known as the equivalence theorem. Noting that the $A_1$, depends on the physical wavelength of the mode with respect to the Hubble momentum, $k/aH$, even if the coefficients of the EEFToI are time independent, {\it i.e.} if $\delta_3\neq -3 \delta_4$, the coefficients of the dispersion relation becomes dependent on the physical momentum of the mode and hence on time. This happens despite the fact that we assumed that the time-dependence of the couplings in the unitary gauge action is slow compared to the Hubble time. Time-independent situation is what we focused on in \cite{Ashoorioon:2018uey} in which we resorted to numerics to analyze the equation of motion. In this work, however, I focus on the more general case, where the coefficients of the dispersion relation are time-dependent.

Expressing the equation of motion for $u_k=a\,\pi_k$, in conformal time and again in the limit $\dot{H}\to 0$, we obtain
\beqa
&&u_k''+ \frac{2 k^2  H^3 (\delta_3+3 \delta_4 )}{a A_1}\, u_k'+u_k \left(\frac{C_1}{A_1}\frac{k^6}{a^4}+\frac{D_1}{A_1}\frac{k^4}{a^2}+\frac{F_1}{A_1}k^2-\frac{a''}{a}\right)=0\,.
\eeqa
 the above equation in a de Sitter space, where $a=-1/(H\tau)$, takes the form
\beqa\label{equ}
u_k''&+& \frac{G_3 \tau k^2 \delta_4}{G_1+G_2 \tau^2 k^2}\,u_k'+u_k
\left(\frac{F_2}{G_1+G_2 \tau^2 k^2 }k^2+\frac{D_2 k^2}{G_1+G_2 k^2 \tau^2}k^2 \tau^2\right.\nonumber\\&+&\left.\frac{C_2 k^2}{G_1+G_2\tau^2 k^2} k^4\tau^4-\frac{2}{\tau^2} \right)=0
\eeqa
where
\beqa\label{var2}
G_1&\equiv&A_0+9 H^4 \left(\delta_3-\frac{3}{2}\delta_4\right)-9\bar M_4 H^3\,, \qquad F_2\equiv F_1\nonumber\\
G_2&\equiv& - H^4 \left(\delta_3+3\delta_4\right)\,,~\quad \qquad \qquad \qquad \qquad D_2\equiv D_1 H^2\nonumber\\
G_3&\equiv&- 2 H^4 \left(\delta_3+3\delta_4\right)=2 G_2\,, \qquad \qquad \qquad C_2\equiv C_1 H^4 
\eeqa
and
\beq
A_1(k,\tau)=G_1+G_2 k^2 \tau^2
\ .
\eeq
Hence all the coefficients including the speed of propagation of the perturbations become time-dependent. In terms of a dimension less variable, $x\equiv k\tau$, the equation of motion \eqref{equ} takes the form
\beq
\frac{d^2 u_k}{d x^2}+\frac{G_3 x}{G_1+G_2 x^2}\frac{d u_k}{d x}+\left(\frac{F_2}{G_1+G_2 x^2}+\frac{D_2 x^2}{G_1+G_2 x^2}+\frac{C_2 x^4}{G_1+G_2 x^2}-\frac{2}{x^2}\right) u_k=0
\ .
\label{eom}
\eeq
This is the equation that we will focus on in the rest of this paper. 
At high momenta energy scales as $k^3$, and thus the mixing energy is
 \beq
E> E_{\rm mix}\sim \frac{|A_1(k/aH)| \sqrt{C_1 }}{M_{\rm Pl}^3}
\ ,
\eeq
which is also a time-dependent quantity. Even if at the horizon crossing, the parameters of the gEEFToI in the unitary gauge are arranged to be such that $ E_{\rm mix}<H$, when the wavelengths of the modes are much smaller than the Planck length, or {\it equivalently} when
 \beq
 \frac{k}{a}\gtrsim M_{\rm Pl} \left(\frac{M_{\rm Pl}}{H}\right)^{1/2}
 \eeq
 the mixing energy becomes larger than the  Hubble parameter during inflation. Hence each individual mode is coupled to gravity at super-Planckian momenta\footnote{$H<10^{-5}~M_{\rm Pl} $ from observation.}. Of course, the Einstein-Hilbert action already breaks down at Planckian distances, and so the formalism of effective field theory becomes applicable only well after this epoch. The length and scale of inflation could be of course such that the CMB modes never undergo such a phase. 
 
 Another impeding factor in following the perturbations  back in time, when $\delta_3\neq -3\delta_4$, is that  the coefficients of the dispersion relation and in particular the speed of sound, would depend on the physical momentum of the mode and when the physical momentum of the mode was large, the quantity that plays the role of speed of sound  would get smaller. This is true even if the parameters are arranged to be such that $c_{{}_S}^2$ in the IR limit, which is roughly given by $F_2/G_1$, is $\mathcal{O}(1)$. In this work we focus on the region of parameter space that the speed of cosmological perturbations in the IR is subluminal, {\it i.e.} $0\leq c_{{}_S}^2 \leq 1$.  For small speed of sound the self interactions of the Goldstone boson enhances and it becomes strongly coupled at certain energy scales such that the unitarity is lost above some threshold energy. For the  $(1+g^{00})^2$ operator, this cutoff is estimated in  \cite{Cheung:2007st} to be 
 \beq\label{Lambda}
 \Lambda^4\simeq 16\pi^2 M_{\rm Pl}^2 \dot{H}\frac{c_{{}_S}^5}{1-c_{{}_S}^2}\,.
 \eeq
In order to be able to use the EFT formalism for the relevant mode, $\Lambda\gtrsim H$. Since now $c_S$ depends on time, $\Lambda$ will be a function of time too. The unitarity is lost when 
\beq
\Lambda(k \tau_u)\simeq H\,.
\eeq 
In order for our analysis below to make sense, one has to assume that $\tau_u$ for each mode, $k$, happens when the mode is well within the regime of domination of $k^6$ term in the dispersion relation. This can be arranged by tuning the parameters of the gEEFToI. Defining $x_u\equiv k\tau_u$, the hypersurface at which the EFT formalism breaks down could be computed to be
\beq
x_u^5\simeq 16\pi^2 \frac{M_p^2}{H^2}\epsilon  \left(\frac{F_2}{G_2}\right)^{5/2}\,,
\eeq
where above  we have assumed that $G_2 x_u^2\gg G_1$. To be in the regime of domination of $k^6$ at $x_u$, 
\beqa
C_2 x_u^4 &\gg& F_2\,,\\
C_2 x_u^2 &\gg& D_2\,.
\eeqa
In the following we have assumed that the above two conditions are satisfied in the gEEFToI formalism.

Various scenarios can arise depending on the relative sign of the parameters $G_1$ and $G_2$. In this work, we assume that the speed of sound and the coefficient of the sixth order correction to the dispersion relation in  IR are positive, {\it i.e.}
\beq
\sign(F_2 G_1)>0\,,\qquad \sign(C_2 G_1)>0.
\eeq
These two conditions are, respectively, necessary to guarantee the stability in the IR and UV limits. The coefficient of the quartic part of the dispersion relation can be positive or negative in the IR, similar to what we analyzed in \cite{Ashoorioon:2017toq,Ashoorioon:2018uey}. 

\section{Analysis of the Mode Equation in gEEFToI}

\subsection{Conversion of the EOM to Confluent Heun Equation}
 
Despite having a complicated form, interestingly, one can obtain an exact solution for the above equation of motion \eqref{eom}. In the rest of analysis, we assume $C_2 G_2>0$, {\it i.e.} these parameters have the same sign.  Plugging the ansatz,
\beq\label{ext-mode}
u_k(x)=x^{-1} e^{-\frac{i}{2}\sqrt{\frac{C_2}{G_2}} x^2} w_k \left(-\frac{G_2 x^2}{G_1}\right),
\eeq
in the above equation, one arrives at the following differential equation for $w_k$
\beq\label{HC-deq-NSCF}
\frac{d^2 w_k(z)}{dz^2}+\left(4p+\frac{\gamma}{z}+\frac{\delta}{z-1}\right) \frac{d w_k(z)}{dz}+\frac{4p\alpha z-\sigma}{z(z-1)} w_k (z)=0\,,
\eeq
where
\beqa
z&\equiv& -\frac{G_2 x^2}{G_1}\,,\qquad p=i \frac{G_1}{4}\sqrt{\frac{C_2}{G_2^3}}\,, \qquad \gamma=-\frac{1}{2},\, \qquad \delta=1\,,\nonumber\\
\alpha&=&\frac{1}{4}-\frac{i}{4}\frac{C_2 G_1-D_2 G_2}{\sqrt{G_2^3 C_2}}\,,\quad \sigma=-\frac{i G_1}{4}\sqrt{\frac{C_2} {G_2^3}}+\frac{1}{2}-\frac{F_2}{4G_2}\, .\label{relation-par-EEFToI-Heun}
\eeqa
 This is nothing other than the ``{\it non-symmetrical canonical form}'' of the  differential equation for Confluent Heun functions \cite{Ronveaux95}. The differential equation has two regular singular points at $z=0,1$ and an irregular singular point at $z=\infty$. The general solution to the differential equation \eqref{HC-deq-NSCF} is a linear combination of the angular (or Frobenius) solution, ${\rm HeunC}^{a}(p,\alpha,\gamma,\delta,\sigma; z)$, normalized as
\beq
{\rm HeunC}^{(a)}(p,\alpha,\gamma,\delta,\sigma; 0)=1\,,
\eeq
and the radial solution, ${\rm HeunC}^{(r)}(p,\alpha,\gamma,\delta,\sigma; 0)$, which is defined by its behavior at $z\to -\infty$
\beq
\lim_{z\to -\infty} z^{\alpha} {\rm HeunC}^{(r)}(p,\alpha,\gamma,\delta,\sigma; 0)=1\,.
\eeq
For complex $p=|p| e^{i\phi}$, the ray ${\rm arg} z=\pi-\phi$ should be taken. One should also note that the argument of the function, $z$, in the case where $\sign(G_1G_2)>0$ is negative and otherwise positive.  In the former case, $-\infty< z \leq 0 $, whereas in the latter one, $0\leq z <\infty$, which includes the singular point $z=1$. As we will see, the regularity of the mode function at $z=1$ could be still satisfied in certain regions of parameter space,  otherwise the mode function and the two point function diverge.

Usually the general solution to the CH differential equation is written as
\beq
w_k(z)= C_1(k) {\rm HeunC}^{(a)}(p,\alpha,\gamma,\delta,\sigma; z)+C_2 (k) z^{1-\gamma} {\rm HeunC}^{(r)}(p,\alpha,\gamma,\delta,\sigma; z)\,.
\eeq
Noting that the prefactor, $\exp\left(-\frac{i}{2}\sqrt{\frac{C_2}{G_2}} x^2\right)$, in eq. \eqref{ext-mode}, is singular when $G_2=0$, or identically when $\delta_3=-3\delta_4$ the explicit solution in \eqref{ext-mode}, becomes singular and meaningless. The corresponding sector, $\delta_3=-3\delta_4$ defines the  Extended Effective Field Theory of Inflation (EEFToI) of inflation \cite{Ashoorioon:2018uey} and we used numerical integration to find solutions to the mode equation. One should also note that the argument of the function in the case where $\sign(G_1G_2)>0$ is negative and otherwise positive.  
On the other hand, the asymptotic forms of the $w_k(z)$ in the neighborhood of the irregular singular point at $z=\infty$ is
\beqa
w_{k}^{1\infty} (z)=z^{-\alpha} \sum_{i=0}^{\infty} a_{i}^{1\infty} z^{-i} \\
w_{k}^{2\infty} (z)=z^{\alpha-\gamma-\delta} \exp(-4pz)\sum_{i=0}^{\infty} a_{i}^{2\infty} z^{-i}\,.
 \eeqa
 Using the above asymptotic forms one can obtain the asymptotic form of the the mode function $u_k(x)$, which becomes
 \beq
 u_k^{\infty}(x)\simeq D_1(k)~e^{-i \frac{x^2}{2} \sqrt{\frac{C_2}{G_2}}}~x^{-\frac{3}{2}+\frac{i}{2}\frac{C_2 G_1-D_2 G_2}{\sqrt{G_2^3 C_2}}}+D_2(k)~e^{i \frac{x^2}{2} \sqrt{\frac{C_2}{G_2}}}~x^{-\frac{5}{2}+\frac{i}{2}\frac{C_2 G_1-D_2 G_2}{\sqrt{G_2^3 C_2}}}\,,
 \eeq
 The asymptotic form of the solution thus contains both positive and negative frequency WKB modes. Starting from WKB positive frequency mode corresponds to choosing $D_1(k)=0$, as one can easily verifies. To know which combination of the HC functions can yield a positive frequency WKB mode in the past infinity, one has to bring the differential equation \eqref{HC-deq-NSCF} to the Leaver form \cite{Leaver},
\beq\label{Leaver-HC-deq}
z(z-1) \frac{d^2 v(z)}{d z^2}+(B_1+B_2 z) \frac{d v(z)}{ d z}+[B_3-2\eta \omega (z-1)+\omega^2 z (z-1)] v(z)=0\, .
\eeq
This could be achieved by the following {\it homotopy} transformation
\beq\label{wv}
w(z)=e^{-2zp} v(z),
\eeq
where
\beqa\label{Heun-EEFT-rel}
B_1=-\gamma=\frac{1}{2}\,, \quad && B_2=\gamma+\delta=\frac{1}{2}\,, \quad B_3=4\alpha p-4\gamma p-2\delta p+\sigma \nonumber\\
 \omega=2i p\,,&&\qquad \eta=i (\alpha-\frac{\gamma+\delta}{2})=i(\alpha-\frac{1}{4})
\eeqa
The factor $e^{-2zp}$ in \eqref{wv} cancels the factor $e^{-\frac{i}{2}\sqrt{\frac{C_2}{G_2}} x^2}$ in \eqref{ext-mode} and thus the mode function will finally take the form
\beq
u_k(x)=x^{-1} v(z)\,.
\eeq

 \subsection{The Case with Singularity in Evolution of the Coefficients, $\mathbf{\mathrm{\sign(G_1 G_2)<0}}$}
 
 \begin{figure}[htbp]
\begin{center}
\includegraphics[scale=0.8]{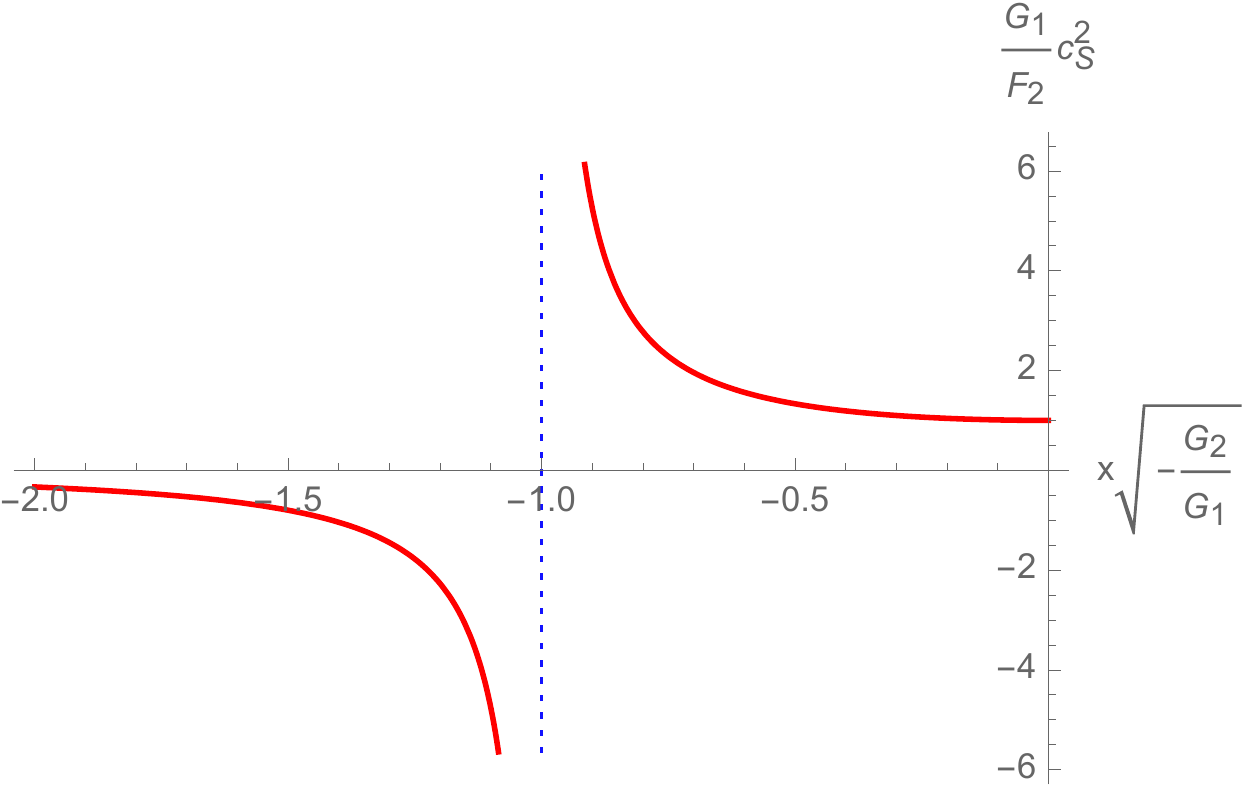}
\caption{The behavior of the the $c_S^2$, as a typical time-dependent coefficient in the dispersion relation, as a function of time when $\sign(G_1 G_2)<0$. There is a singularity in time and a change in sign of the coefficients at the singularity.  When the physical wavelength of the mode is very small, the speed of sound (and other coefficients of the dispersion relation) go to zero. When $\Lambda~H$  for each mode, there will be a loss of unitarity for each mode. The parameters in the problem are tuned such that this loss of unitarity happens well before the transition from $\omega^2\sim k^6$  to the $\omega^2\sim k^4$ which guarantees that the mode spends a substantial amount during the sextic dispersion relation and make it relevant.}
\label{sing}
\end{center}
\end{figure}

In this case $0\leq z<\infty$ as $x$ runs from $0$ to $-\infty$. Even if $c_S^2$ and the coefficient of the sixth order term, $k^6$, in the dispersion relation are assumed to be positive in the IR, there is a moment in the evolution of each mode when they become singular and  negative for the times before the singularity point, please see figure \ref{sing}, where we have only plotted $c_S^2$ as a function of variable that is roughly proportional to time. The coefficient of the quartic term in the dispersion relation could be positive or negative in the IR, but also becomes singular and changes sign for earlier times. In fig. \ref{sing}, we have only plotted the speed of sound, $c_S^2$, as a function of $x$ and assumed to be positive in the limit $x\to 0$. The asymptotic form of the solution at $z\to \infty$  can be obtained from the normal Thom\'e solutions \cite{Leaver} to be, in general, a combination of
\beq
\lim_{z\to \infty} v(z) \sim \mathrm{e}^{\pm i \omega z} z^{\mp i \eta-B_2/2} \,.
\eeq
The Jaff\'{e} solution that behaves like the positive frequency WKB mode in infinite past is
\beq\label{Jafex>1}
v_{\rm J} (z)= e^{i\omega z} z^{-B_2/2-i \eta }\sum_{n=0}^{\infty} a_n \left(\frac{z-1}{z}\right)^n
\eeq
The solution is valid in the interval $1\leq z< \infty$. The coefficients obey the three term recurrence relation
\beqa\label{rec-rel}
&&\alpha_0 a_1+\beta_0 a_0=0\,,\nonumber\\
&&\alpha_n a_{n+1}+\beta_n a_n+\gamma_n a_{n-1}=0, \qquad n=1,2,\ldots
\eeqa
where
\beqa\label{Coeff-series}
\A_n&=&(n+1) (n+B_2+B_1/z_0)=(n+1)^2\,,\\
 \B_n&=&-2n^2-2 \left[B_2+i(\eta-\omega z_0)+B_1/z_0\right]n-(B_2/2+i\eta) (B_2+B_1/z_0)\nonumber\\
 &+&i\omega (B_1+B_2 z_0)+B_3  \nonumber\\
&=& -2n^2-2[1+i(\eta-\omega)]n-(1/4+i\eta)+i\omega+B_3\\
 \G_n &=& (n-1+B_2/2+i\eta) (n+B_2/2+i\eta+B_1/z_0) =(n+i\eta)^2-9/16
\eeqa
The recurrence relation \eqref{rec-rel} have two set of solutions ${a_n^{(1)}: n=1,2,\ldots}$ and $\{a_n^{(2)}: n=1,2,\ldots\}$ where where $\lim_{n\to\infty}a_n^{(1)}/a_n^{(2)}=0$. Then the solution which is identified by $a_n^{(1)}$ is called {\it minimal} and the other one, with $a_n^{(2)}$ coefficients, is referred to a {\it non-minimal} or {\it dominant} solution. As noted earlier for the case where $\sign(G_1G_2)>0$, $z$ runs from $-\infty$ to $0$ which avoid the singular point $z=1$, whereas in the case where $\sign(G_1G_2)<0$, the mode function passes through the singular point $z=1$.  In this case, the eigensolutions of the Confluent Heun equation are the ones for which $\sum_{n=0}^{\infty}a_n$ converges \cite{Leaver}. Then the solution at $z\to \infty$ has a pure  $e^{i\omega z} z^{-B_2/2-i\eta}$
behavior at $z\to \infty$ and is regular at $z=1$. The sum $\sum_{n=0}^{\infty}a_n$ usually converges iff the $a_n$'s are the {\it minimal} solution to the recurrence relation. For the minimal solution, the ratio of successive elements in the large $n$ limit goes to zero. It has been proven in \cite{Gautschi} that the ratio of successive coefficient of the minimal solution is given by the continued fraction
\beq
\frac{a_{n+1}}{a_n}=\frac{-\G_{n+1}}{\B_{n+1}-}\frac{\A_{n+1}\G_{n+2}}{\B_{n+2}-}\frac{a_{n+2}\G_{n+3}}{\B_{n+3}-}\,.
\eeq
For $n=0$ this yields
\beq
\frac{a_1}{a_0}=\frac{-\G_{1}}{\B_{1}-}\frac{\A_{1}\G_{2}}{\B_{2}-}\frac{a_{2}\G_{3}}{\B_{3}-}\,,
\eeq
which along with the first of Eqs. \eqref{rec-rel}
\beq
\frac{a_1}{a_0}=-\frac{\beta_0}{\alpha_0}\,,
\eeq
requires that
\beq\label{cont-frac}
0=\beta_0-\frac{\A_0\G_1}{\B_1-}\frac{\A_1\G_2}{\B_2-}\frac{\A_2\G_3}{\B_3-}\cdots\,.
\eeq
Since $\A_n$, $\B_n$ and $\G_n$ are functions of $B_3$, $\omega$ and $\eta$, which are related to parameters in the gEEFToI through Eqs. \eqref{relation-par-EEFToI-Heun} \& \eqref{Heun-EEFT-rel}, this sets a relation between the gEEFToI parameters that should be satisfied to make the mode function regular at $z=1$. In other words, when $\sign(G_1 G_2)<0$, not for all parameter in the gEEFToI the mode function is regular at $z=1$.


Even if, arranging for the parameters to satisfy the relations \eqref{cont-frac}, the mode function does not become singular at the point $z=1$, the power spectrum does not seem to be finite and well defined in the limit $x\to 0$. The power spectrum is defined as
\beqa\label{power-spec}
P_S&=&{\left(\frac{k^3}{2\pi^2}\right)}^{1/2}\left|\frac{u_k}{a}\right|_{k\tau\to 0}=\lim_{x\to 0}2 x^2 |u_k|^2=\lim_{x\to 0} 2 |v_k|^2 \\
&\propto&\lim_{x\to 0} x^{-1-\frac{C_2 G_1-D_2 G_2}{\sqrt{G_2^3 C_2}}} \left|\sum_{n=0} a_n \left(-\frac{g_1}{g_2}\right)^n \frac{1}{x^{2n}} \right|^2\,.
\eeqa
The above expression goes to infinity in the limit of $x\to 0$. The mode function has an {\it essential singularity} at $x=0$ which makes the power spectrum, for general parameters, infinite. One might be tempted to think that the parameters could be tuned to terminate the series at some value of $n_{\ast}$ which would yield an effective power-law format to the power spectrum. For that to happen, {\it i.e. $a_{\nst+1}=0$}, $\beta_{\nst}$ and $\gamma_{\nst}$ should be both set to zero. By adjusting the parameters, this could be in principle achieved. However to terminate the series $a_{\nst+2}$ should be set to zero which, in turn, requires that $\G_{\nst+1}=0$. Looking at the third equation of \eqref{Coeff-series}, this cannot be achieved for both $n_{\ast}$ and $n_{\ast}+1$. Thus the possibility of the series getting terminated is also out of question. This shows that if one assumes that the mode has started from the vacuum, which corresponds to the positive frequency WKB mode, the power spectrum turns out to be infinite!


In a more realistic situation, one expects that there will be a beginning to inflation, corresponding to initial conformal time $\tau_i$. If for the given mode $k$,
\beq\label{avoid-sing}
z_i\equiv -\frac{G_2 x_i^2}{G_1}<1,
\eeq
where $x_i\equiv k\tau_i$, instead of the Jafe solution \eqref{Jafex>1}, which is valid in the regime of $z_0\leq z<\infty$ and is only extendable to the region of $0<z<z_0$ if the eigensolution condition is satisfied, one can instead use Baber and Hasse solution which is convergent in the region $0\leq z< z_0$. Their solution is given as
\beq
v_{\rm B-H}=\exp(i\omega z) \sum_{n=0}^{\infty} a'_n z^n\,,
\eeq
The sequence of expansion coefficients are given be the three term recurrence relation
\beqa
&&\alpha'_0 a'_1+\beta'_0 a'_0=0\,,\nonumber\\
&&\alpha'_n a'_{n+1}+\beta'_n a'_n+\gamma'_n a'_{n-1}=0, \qquad n=1,2,\ldots
\eeqa
where
\beqa
\A'_n&=&-n^2+(B_1-1) n+B_1 \,,\\
 \B'_n&=& n^2+(B_2-2i\omega -1) n+2\eta \omega +i\omega B_1+B_3\,\\
 \G'_n &=& 2i \omega n +i\omega (B_2-2)-2\eta \omega
\eeqa
In this case when $x\to 0$, the mode function will approach a constant value $a'_0$ and thus the power spectrum becomes constant. For this to happen nonetheless, it is necessary that the condition \eqref{avoid-sing} is satisfied for all the modes that exit inflation during the $N_e$ that exit during inflation. Condition \eqref{avoid-sing} is scale-dependent through the variable $x_i=k\tau_i$ and if it holds for the largest $k$ that exit during inflation, {\it i.e.} $k=H$, it is naturally satisfied for smaller $k$'s too. This will lead to the following constraint on the parameters of the gEEFToI in this case
\beq
-\frac{G_2 H^2 \tau_i^2}{G_1}<1\,,
\eeq
or equivalently if
\beq
e^{2N_i}<-\frac{G_1}{G_2}\,.
\eeq
Of course, oting that $N_i$ has to be bigger than the number of e-folds needed to solve the problems of the Big Bang cosmology, $N_i\gtrsim 60$, this condition is quite restrictive and indicates a hierarchical separation between the parameters $G_1$ and $G_2$. Nonetheless, arranging for this,  it seems to be possible to have a well-defined finite power spectrum in the infinite IR.

%

\subsection{Case of positive definite sound speed squared, $\mathrm{\sign(G_1 G_2)>0}$}

 \begin{figure}[htbp]
\begin{center}
\includegraphics[scale=0.8]{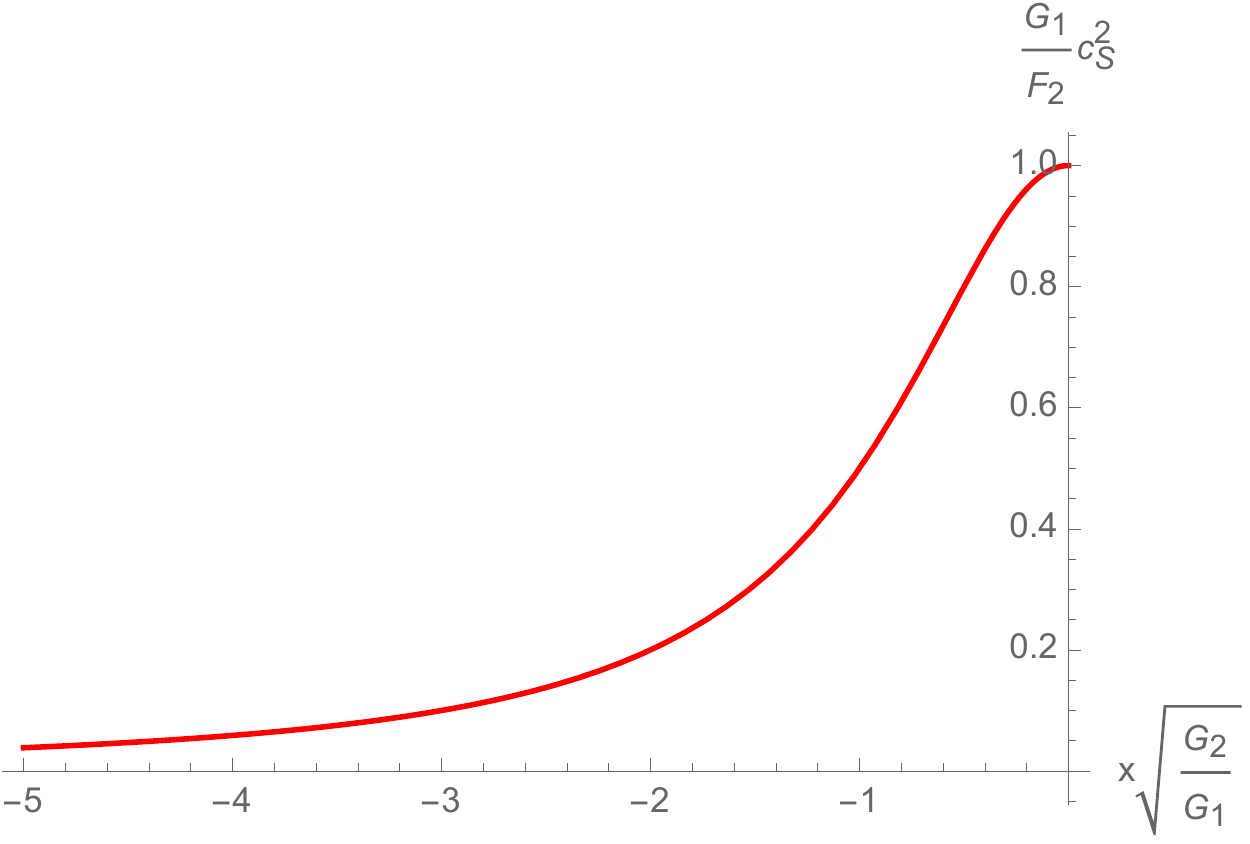}
\caption{The behavior of the the $c_S^2$, as a typical time-dependent coefficient in the dispersion relation, as a function of time when $\sign(G_1 G_2)>0$.  In this case the speed of sound, and other coefficients of the dispersion relation, approach zero and a constant value in infinite past and future. Contrary to the previous case, there will be no singularity in the evolution of the mode. However as in the previous case, when $\Lambda~H$,  there will be a loss of unitarity for each mode.}
\label{nonsing}
\end{center}
\end{figure}

In this case, $z\in (-\infty,0]$ as $x$ runs from $+\infty$ to $0$. The regular singularity at $z_0=1$ is out of $ (-\infty,0]$ interval in this case. It can be shown that the the behavior of the solution in the region $z\in (-\infty,0]$ could be mapped to the region $[z_0,\infty)$, where in our case $z_0=1$. To see this, let us do the following reparameterization
\beq
z'\equiv z_0-z\,,
\eeq
which leaves the form of confluent Heun function equation, \eqref{Leaver-HC-deq}, intact with new parameters that are defined as follows
\beqa
B_1'=-(B_1+B_2 z_0)=-1\,&&\qquad B_2'= B_2=1/2\,, \qquad B_3'= B_3+2\eta \omega z_0 \\
\eta'=-\eta\, &&\qquad \omega'=\omega
\eeqa
The Jaff\'{e} solution that resembles the positive frequency WKB mode is given as
\beq\label{vprime}
v'_J(z')=\exp(-i \omega z') z'^{-B_2'+i \eta'} \sum_{n=0}^{\infty} b_n\left(\frac{z'-z_0}{z'}\right)^n
\eeq
The expansion coefficients are produced by the three-term recurrence relation,
\beqa
\tilde{\A}_0 b_1+\tilde{\B}_0 b_0=0\,,\\
\tilde{\A}_n b_{n+1}+\tilde{\B}_n b_n+\tilde{\G}_n b_{n-1}=0\,.
\eeqa
where the recurrence coefficients are given by
\beqa
\tilde{\A}_n&=&(n+1) (n+B_2'+B_1'/z_0)\,\\
\tilde{\B}_n&=&-2 n^2-2 [B_2-i(\eta'-\omega' z_0)+B_1'/z_0] n-\nonumber \\&& (B_2'/2-i \eta')(B_2'+B_1'/z_0)-i\omega (B_1'+B_2' z_0)+B_3'\\
\tilde{\G}_n&=& (n-1+B_2'/2-i\eta) (n+B_2'/2-i\eta'+B'_1/z_0)
\eeqa
The power spectrum, as defined in the first line of \eqref{power-spec}, turns out to be
\beq
P_S=\lim_{x\to 0} 2x^2 |u_k(x)|^2= \lim_{x\to 0} 2x^2 |x^{-1} v'_J(z')|^2=\lim_{z'\to z_0} |v'_J(z')|^2=2 |b_0|^2\,,
\eeq
which is finite. On the other hand the finiteness of  $v'_J(z')$ at infinity $z'\to \infty$, requires that $\mathcal{B}\equiv\sum_{n=0}^{\infty} b_n$ should converges which, in turn, means that the solution \eqref{vprime}, should be the {\it eigensolution} of the mode equation. As stated in the previous section the sum over $b_n$'s converge if and only if they are the minimal solution to the recursive equation, which corresponds to satisfying the continued fraction equation,
\beq
\frac{b_{n+1}}{b_n}=-\frac{\tilde{\G}_{n+1}}{\tilde{\B}_{n+1}-}\frac{\tilde{\A}_{n+1}\tilde{\G}_{n+2}}{\tilde{\B}_{n+2}-}\frac{\tilde{\A}_{n+2}\tilde{\G}_{n+2}}{\tilde{\B}_{n+3}-}
\eeq
which requires $\omega$ to be the root of
\beq\label{cont-frac-omega}
0=\tilde{\B}_0-\frac{\tilde{\A}_0\G_1}{\B_1-}\frac{\tilde{\A}_1\tilde{\G}_2}{\tilde{\B}_2-}\frac{\tilde{\A}_2\tilde{\G}_3}{\tilde{\B}_3-}\,.
\eeq
The above equation is only valid for certain characteristic values of $\omega$. Since the recurrence coefficients are functions of $\omega$, $\eta$ and $\sigma$, this will provide constraints on the parameters of the dispersion relation for which the resulted power spectrum in the IR limit is finite and well-defined. If the parameters do not satisfy these constraints, the power spectrum is indefinite in the large wavelength limit. 

The constraint \eqref{cont-frac-omega}, could also be obtained from the Wronskian condition which fixes the normalization of the mode function. In principle the normalization of the mode function, determined by $b_0$, could be determined by the Wronskian condition,
\beq\label{Wronsk-u}
k\mathcal{W}(u_k(x),u^{\ast}_k(x))=\frac{i}{\kappa(x)}\,,
\eeq
where
\beq
\kappa(x)=\exp(-\int \frac{2 G_2 x}{G_1+G_2 x^2})=(G_1+G_2 x^2 )\,.
\eeq
The above in terms of the Wronskian of $v'_J(z')$ could be written as
\beq\label{Wronsk-v'}
\mathcal{W}(v'_J(z'),v'^{\ast}_J(z'))=\frac{i\sqrt{z'-1}}{z'}\,.
\eeq
Since we do not know the explicit $n$ dependence of the expansion coefficients, $b_n$'s, summing up the L.H.S. of the equation  \eqref{Wronsk-v'} is a formidable task. If we look at the behavior of $v'_J(z')$ in the limit of $z'\to \infty$, $v'(z')\propto \exp(-i(\omega z'-\eta \ln z'))$, we see that the Wronskian reduces to
\beq\label{b0-wronsk}
\mathcal{W}(v'_J(z'),v'^{\ast}_J(z'))=|\mathcal{B}|^2 \frac{2i \omega}{\sqrt{z'}}\simeq  \frac{i}{\sqrt{z'}}\,.
\eeq
From the three-term recurrence relation all the expansion coefficients $b_n$ could be expressed in terms of $b_0$ and, hence, relation \eqref{b0-wronsk} can use to constrain the value of $b_0$, which in turn determine the amplitude of the power spectrum. In the first approximation that $B\sim b_0$, the Wronskian condition \eqref{b0-wronsk} tells us that the normalization
\beq
|\mathcal{B}|\sim \frac{1}{\sqrt{2\omega}}
\eeq
which is nothing other than Bunch-Davies vacuum normalization. Of course, we will have higher order corrections to the frequency of the mode from the unharmonic nature of the mode equation. Still the value of $b_0$ obtained from eq. \eqref{b0-wronsk} should be finite which, in turn, translates to the convergence of  $\mathcal{B}$.

\section{Excited Initial Condition and Finite Two Point Function in gEEFToI}

There is a one-side series expansion for the Confluent Heun functions in terms of Confluent Hypergeometric functions \cite{Leaver,CHE-JMP} which converges for any $z$ and is regular at both $z=z_0$ and $z=0$ \footnote{Since, neither $B_2+B_1/z_0$ nor $i \eta+\frac{B_2}{2} $ are zero or negative integers, the solution could be written as one-sided series, running from $n=0$ to $\infty$.},
\beq\label{Heun-KummerU}
v_1(z)=e^{-i\omega z} \sum_{n=0}^{\infty} \frac{(-1)^n c_n} {\Gamma(n+B_2)} \Phi\left(\frac{B_2}{2}-i\eta, n+B_2;2i\omega z\right)\,.
\eeq
The recurrence relation for the expansion coefficients are
\beqa
\hat{\A}_0 c_{1}+\hat{\B}_0 c_0=0\\
\hat{\A}_n c_{n+1}+\hat{\B}_n c_n+\hat{\G}_n c_{n-1}=0
\eeqa
where the recursion coefficients are
\beqa
\hat{\alpha}_n&=&(n+1)(n+i\eta+B_2)\, \nonumber\\
\hat{\beta}_n &=& n(n+B-2-1+2i\omega z_0)+B-3+i\omega z_0 \left(B-2+\frac{B_1}{z_0}\right)\, \nonumber\\
\hat{\gamma}_n &=& 2 i \omega z_0 (n+B_2+\frac{B_1}{z_0}-1)\, \nonumber\\
\eeqa
where $\Phi(a,c;y)$ is the regular confluent hyper-geometric function, which is the solution to the differential equation
\beq
y \frac{d^2 \varphi}{d y^2}+(c-y)\frac{d \varphi}{d y}-a\varphi=0\,,
\eeq
and defined as
\beq
\Phi(a,b,z)=\sum_{n=0}^{\infty} \frac{(a)_n z^n}{(b)_n}\,,
\eeq
where $(a)_n$ is the Pochammer's symbol defined as
\beq\label{CHF-eq}
(a)_n\equiv a(a+1)(a+2)\cdots(a+n-1)\,, \qquad {\rm and}\qquad (a)_0\equiv 1\,.
\eeq
According to \cite{CHE-JMP}, the above solution converges for any value of $z$ and since $v_1(z)\sim {\rm Const.}$, for $z=0$, the power spectrum will be finite in the case where the speed of sound changes sign and becomes singular at some point during the evolution. In the case where $c_S^2$ is positive definite and finite during the evolution, the power spectrum is finite  too.

There are two other series expansions in terms of hypergeometric functions that have a purely negative and positive frequency WKB mode in infinite past
\beqa
v^{\infty}&=&e^{-i\omega z}\sum_{n=0}^{\infty} (-1)^n c_n \Psi\left(\frac{B_2}{2}-i\eta, n+B_2;2i \omega z\right)\nonumber\\
\check{v}^{\infty}&=&e^{i\omega z}\sum_{n=0}^{\infty}  d_n \Psi\left(n+i\eta+\frac{B_2}{2}, n+B_2;-2i \omega z\right)\nonumber
\eeqa
The coefficients $d_n$ is also determined recursively through the following relations,
\beqa\label{heun-kummerU}
{\check{\alpha}}_0 d_{1} +{\check{\beta}}_0 d_0&=&0\,,\\
\check{\alpha}_n d_{n+1} +\check{\beta}_n d_n+ \check{\gamma}_n d_{n-1}&=&0\,,
\eeqa
where
\beqa
\check{\A}_n&=&n+1\,,\\
\check{\B}_n&=& n(n+B_2-1+2 i \omega z_0)+B_3+i\omega z_0 \left(B_2+B_1/z_0\right)\,,\\
\check{\G}_n&=&2i \omega z_0 \left(n+B_2+\frac{B_1}{z_0}-1\right)\left(n+i\eta+\frac{B_2}{2}-1\right)\,.
\eeqa
$\Psi(a,c;y)$ now is the irregular confluent hypergeometric function which is the solution to the differential equation \eqref{CHF-eq} and in the case where  $a$, $c$ and $c-a$ are not integers, which is true in the above cases, is related to the regular solution, $\Phi(a,c;y)$, through the  relation
\beq
\Psi(a,c;y)=\frac{\Gamma(1-c)}{\Gamma(a-c+1)} \Phi(a,c;y)+\frac{\Gamma(c-1)}{\Gamma(a)} y^{1-c} \Phi (a-c+1,2-c;y)
\eeq
The two solutions, $v^{\infty}$ and $\check{v}^{\infty}$ converge only for $|z|>|z_0|=1$. Their asymptotic forms at $z\to \infty$, respectively, are
\beqa
\lim_{z\to\infty} v^{\infty}(z)&\sim& \exp(-i\omega z) z^{i\eta-\frac{B_2}{2}}\,,\\
\lim_{z\to\infty} \check{v}^{\infty}(z)&\sim& \exp(i\omega z) z^{-i\eta-\frac{B_2}{2}}\,.
\eeqa
Since $v^{\infty}(z)$, $\check{v}^{\infty}(z)$ and $v_1(z)$ are three solutions to a linear second order differential equation, it is  possible to express one of the solutions in terms of the other two, in the region that all three are viable, {\it i.e.} $|z|>1$. In particular in the region of validity of $v^{\infty}$ and $\check{v}^{\infty}$, namely $|z|>1$, one can express $v_1(z)$ in terms of $v^{\infty}(z)$ and $\check{v}^{\infty}(z)$, which respectively behave like positive and negative frequency WKB modes at $z\to \infty$ (or in the latter case $z'\to\infty$). This means that the solution which would yield a finite two-point function, irrespective of the behavior of the value of the coefficients, emanates from an excited state. It is interesting that the details of the initial condition reflects itself in whether the two-point function asymptotes to a constant value. Only for a specific excited initial condition the power spectrum turns out to be finite in the IR limit. For any other state other than the one which reduces to the form $v_1(z)$, the power spectrum will be divergent.

The reason for the finiteness of the power spectrum for all values of the parameters could be traced back to the fact that for each mode, at sufficiently small physical wavelength, the unitarity is lost, even if it is arranged to be satisfied at the Hubble crossing. This loss of unitarity comes about because the corresponding speed of sound for the perturbations goes to zero, which results in the reduction of the cutoff scale, $\Lambda$, for the gEEFT obtained from non-renormalizable operators. For each mode, the time span where $\Lambda(k\tau)\lesssim H$ is the part of the evolution that is not governed by the EFT. Whatever the result of evolution during the non-unitary phase is, this computation suggests that assuming the mode having remained in the vacuum of the theory after through the evolution in this phase, leads to inconsistencies. These inconsistencies show up in physically infinite results, like the case where there is a singularity in the time-dependent dispersion relation, or a fine-tuned parameter space that yield a finite two-point function. 

The demand for a completion of the theory to a unitary theory at arbitrarily short wavelengths seems to be entangled with the quest for quantum gravity in this setup too, where the mixing momentum between the Goldstone boson and gravity turns out to be a function of time too, 
\beq
 k_{\rm mix}\sim \frac{\sqrt{A_1(k,\tau)}}{M_{\rm Pl}}=\frac{\sqrt{G_1+G_2 x^2}}{M_{\rm Pl}}
\ .
\eeq
If
\beq
\frac{\sqrt{G_1+G_2}}{M_{\rm Pl}}\ll H
\ ,
\eeq
assuming that the dispersion relation is dominant by the $k^2$ at the horizon-crossing, the Goldstone mode will be decoupled from gravity. However, tracing the mode back in time and at shorter wavelengths, the mixing energy for a given comoving mode with momentum $k$ will increase like $k^6$ and  ultimately will become bigger than $H$. The coupling to gravity becomes important when
\beq
x^2=\frac{M_{\rm Pl}^3 H}{G_2\sqrt{C_1}}\simeq \left(\frac{M_{\rm Pl}}{H}\right)^3
\eeq
where in the second line we have assumed that $\delta_3+\delta_4\sim\delta_3+3\delta_4\sim\mathcal{O}(1)$.  Since $H\lesssim 10^{-5} M_{\rm Pl}$, this suggests that the coupling to gravity happens at sub-Planckian wavelengths. This indicates that the quest for quantum gravity is complementary to the question of the unitary completion of the evolution of the mode at small wavelengths.

\section{Conclusions}
\label{Concl}

I studied the general Extended Effective Field Theory of Inflation (gEEFToI) in which the coefficients of the sixth order polynomial dispersion relation for each mode could become time-dependent. This happens even if the couplings of the spatially diffeomorphism invariant operators in the unitary gauge action are assumed to be independent of time. This is a general situation that occurs if two ghost-free operators, at the order $(\nabla K_{\mu\nu})^2$, are not tuned to satisfy the relation $\delta_3=-3\delta_4$, assumed to hold in \cite{Ashoorioon:2018uey}. Depending on the values of these couplings, the coefficients in the dispersion relation could either become singular and change sign at some point during the evolution of each mode or could monotically increase from a small value in infinite past to a constant value in IR. In both cases, when the mode is smaller than a certain threshold, the unitarity is lost and the Goldstone mode is coupled to gravity, even if the parameters of the unitary gauge action are arranged such that the unitarity is preserved and the Goldstone boson is decoupled from gravity at horizon crossing. However, after the mode gets stretched out of the ``non-unitary phase'' due to the universe expansion, the equations of motion derived within the gEEFToI approach could be still trusted. With the assumption of starting the modes from the vacuum dictated by the gEEFToI equation of motion, which in this case corresponds to the positive frequency WKB mode, the two point function diverges in the infinite IR limit, if the coefficients of the dispersion relation have a singularity and change sign. With the assumption of validity of the WKB mode as an initial condition for the modes, this imposes certain inequalities on the parameters of unitary gauge action in the gEEFToI, which should be respected in order to obtain finite result, {\it i.e.} $\sign(G_1G_2)\nless 0$, or instead the length of inflation should be such that the mode never experiences the region before the singularity. The latter demands a hierarchical relation between two parameters that should be respected. In the other case, where the dispersion relation coefficients asymptote to a constant value, without encountering a singularity, the two point function becomes constant but the requirement that the mode function -- which is expressed as an infinite series expansion-- remains finite at infinite past, dictates that the solution be an eigensolution of the Confluent Heun differential equation. This requires that  some characteristic eigensolution equation given as infinite continued fraction holds which constraints the region of parameters space of the gEEFToI. All these results were under the assumption that the mode function starts from the vacuum solution. 

With the assumption that the mode function and the power spectrum should remain finite, irrespective of the values of the parameters in the gEEFToI, I looked for a solution which is regular in infinite IR limit, for both two scenarios explained above.  I showed that such a mode function is a combination of positive and negative frequency modes in infinite UV limit, {\it i.e.} it is an excited state. This  result is interpreted noting the fact the unitarity is lost at arbitrarily short wavelengths, even though it is arranged to hold at the Horizon crossing. Also the Goldstone mode couples to gravity at sub-Planckian wavelengths. Unless one assumes that the mode emerges as an excited state out of this non-unitary phase, the two-point function for perturbations either turns out to be infinite or finite for a confined part of the parameter. In fact the criterion of finite two-point function can be used to constrain the mode function. Such a mode function turns out to be an excited state. The non-unitary stage of evolution in the gEEFToI, once the mode is stretched to the unitary phase evolution, could be approximated with an excited state. We expect that whatever ``{\it new}'' physics that replaces the non-unitary phase of evolution should produce such an excited mode function. This is constraining for the new physics that should be replaced with the non-unitary stage evolution of the mode. 

         In order to check the unitarity and find the region of parameter space in which the gEEFToI is a fully viable effective field theory, one should compute the three-point function and map out the region of parameter space in which the non-gaussianity does not violate the unitarity bounds.  Noting the current computations at the level of two-point function we have two ways to proceed. Either one starts off from the WKB vacuum for which the parameter space is ``already constrained'' by eq. \eqref{cont-frac-omega} and then compute the bispectrum. The other option would be computing the bispectrum with the mode function \eqref{Heun-KummerU}, which is an excited mode function in the past, and then checking that the level of non-gaussianity does not violate the unitarity bounds. In both cases, in analogy with the work of super-excited states \cite{Ashoorioon:2013eia,Ashoorioon:2017toq} , it is  expected that at least for the region of parameter space where the horizon crossing happens in the regime of domination of the quadratic term in the dispersion relation, $\omega^2\propto k^2$, the generated non-gaussinity would be small and unitarity is under control. We will postpone the investigation of this point in detail to a future study.

As we noted, such new physics, is either due to the nonlinear interactions of the Goldstone mode or the coupling of the Goldstone mode with the gravity at sub-Planckian scales. Being equipped with the preferred mode function, one can study the higher point functions which will then incorporate the signature of new physics further. By studying the correlation functions, one can then learn about the potential new physics signatures in the CMB correlation functions.

\section*{Acknowledgement}

I am thankful to R. Casadio and M. M. Sheikh-Jabbari for useful discussions on this. The author is thankful to the INFN fellowship at the University of Bologna, where this project was initiated, and to the Riemann Center for Geometry and Physics at the Leibniz University in Hannover where it was completed.

\bibliographystyle{JHEP}
\bibliography{bibtex}

\end{document}